\documentstyle[hip-artc]{article}  
\volnumber{}  \edyear{}  \frompage{} \topage{}                           
\recrevdate{November 2000}                                               
\def\rightmark{Do $\bar{Y}$ signal the QGP?}
\def\leftmark{Carsten Greiner}

\def\Journal#1#2#3#4{{#1} {\bf #2}, #3 (#4)}

\def\NPA{{\em Nucl.~Phys.} A}
\def\PLB{{\em Phys.~Lett.}  B}
\def\PRL{\em Phys.~Rev.~Lett.~}
\def\PRD{{\em Phys.~Rev.} D}
\def\PRC{{\em Phys.~Rev.} C}

\def\EPJC{{\em Eur.~Phys.~J.} C}
\def\PRep{\em Phys.~Rep.~}

\newcommand{\expl}{\langle \!\langle}
\newcommand{\expr}{\rangle \!\rangle}

\title{Do chemically saturated antihyperon abundancies
\\[2mm]
signal the quark gluon plasma?}
\authors{
{\twerm Carsten Greiner$^{a,b}$ %
}\\[2.812mm]
{\normalsize
Institut f\"ur Theoretische Physik, Universit\"at Giessen, \\
D-35392 Giessen, Germany
}}
 
\abstract{
We first review the production and the possible chemical
equilibration of strange particles at CERN-SPS
energies within a
microscopic hadronic transport calculation.
It is shown in particular that the strange quarks are
produced initially via string excitations in the primary,
secondary and ternary interactions.
We then further elaborate on a recent idea of antihyperon
production by multi-mesonic reactions like
$n_1\pi + n_2 K \rightarrow \bar{Y}+p $
corresponding to the inverse of the strong binary
baryon-antibaryon annihilation process.
It is argued that by these reactions
the (rare) antihyperons are driven
towards local chemical equilibrium with pions, nucleons and kaons
on a timescale of 1--3 fm/c in the still moderately
baryon-dense initial hadronic environment after the termination of the
prehadronic string phase. Accordingly this mechanism can
provide a convenient explanation for the antihyperon yields
at CERN-SPS energies
without any need of a deconfined quark gluon plasma phase.
}
\keyword{relativsitic heavy ion collisions, quark gluon plasma,
(multi-)strange particles}
\PACS{25.75.-q, 12.38.Mh }
 
\begin{document}
 
\maketitle

\section{Introduction and Motivation}

The prime intention for present and future ultrarelativistic heavy ion collisions
lies in the possible experimental identification of
the quark gluon plasma (QGP). The QGP represents a theoretically hypothesized
and from QCD lattice calculations convincingly established
new phase of matter, where quarks and gluons are deliberated from the
hadronic particles and move freely over an extended, macroscopically large
region. Moreover, considering several different observations
within the Lead Beam Programme at the CERN-SPS,
strong `circumstantial evidence' for the temporal formation of the QGP
has been conjectured \cite{HJ00} very recently.
As one particular example, strangeness enhancement
has been predicted already a long time ago as one of the much favored
diagnostic probes for the short-time existence of a QGP \cite{KMR86}:
The main idea is that the strange (and antistrange) quarks are
thought to be produced more easily and hence also more abundantly
in such a deconfined state as compared to the production via
highly threshold suppressed inelastic hadronic collisions.
In this respect, especially the antihyperons and also the
multistrange baryons were advocated as the
appropriate candidates \cite{KMR86}.

\begin{figure}[htb]
\insertplot{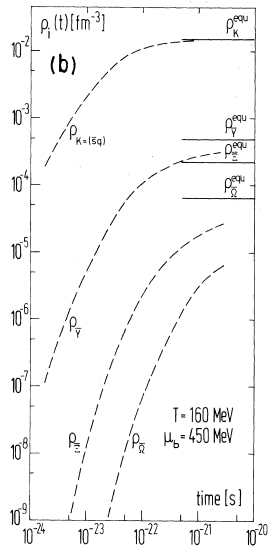}
\caption[]{
Part (b) of Fig. 5.5 taken from the
report article by Koch, M\"uller and Rafelski
\cite{KMR86}. Clearly the antihyperons do not approach
chemical equilibrium even after 1000 fm/c, whereas the kaons have
equilibrated much earlier.
}
\label{fig:KMR}
\end{figure}

In Fig. \ref{fig:KMR} we depict the intriguing observation
from the seminal report article of Koch, M\"uller and Rafelski:
It shows the approach to chemical equilibrium of the various
population densities of strange hadronic particles
containing at least one antistrange quark as a function of time within
a hot and baryonrich hadronic system. Even after 1000 fm/c the antihyperons
do not approach by far their chemical equilibrium values!
It was argued that within a thermalized
fireball environment of hadronic particles
the strange antibaryons are dominantly be produced by
subsequent binary strangeness exchange reactions with the
(maybe) chemically equilibrated kaons like
\begin{equation}
\label{sprodb}
K + \bar{p} \,  \rightarrow \,  \pi + \bar{\Lambda }, \bar{\Sigma }
\, \, \, ; \, \,
K + \bar{\Lambda } \,  \rightarrow \,  \pi + \bar{\Xi }
\end{equation}
with very low cross sections.
On the other hand, assuming the
existence of a temporarily present phase of QGP and
following simple coalescence estimates
the abundant (anti-)strange quarks
can easily combine with the light (anti-)quarks to form the
strange hadrons \cite{KMR86}, which do then, in return, come close to their
chemical equilibrium values.
(Of course, these predictions can only be regarded as qualitative,
yet plausible: A satisfactory theoretical understanding of the
dynamics and of the hadronisation of a hypothetical deconfined phase
as well as the production of strangeness in this state is at
present not really given.)

Indeed, an enhancement of strangeness has been reported,
calibrated in relation to p+p or p+A collisions \cite{SQM98}.
This is  in particular true for the antihyperons (and to a little
lesser extent for multistrange baryons).
Such an enhancement can certainly not be explained by the above mentioned
binary strangeness exchange reactions.
On the theoretical side the analysis of measured
abundancies of hadronic particles
with simple thermal models
\cite{BMS96,CR00,KM00} strongly supports the idea of having established
an equilibrated fireball in some late stage of the reaction,
where all hadrons
with light and/or strange quark content
do exist in number nearly according to their chemical equilibrium values.
(The thermodynamical properties found by the various groups
can phenomenologically easily be explained in fact by a rapidly hadronizing and
disintegrating QGP phase \cite{Sp98}.
This has very recently been pursued in trying to extract
the critical energy density \cite{KM00}.)
(Nearly perfect) Chemical equilibration is found to be true for the
antihyperons.
Alas, this all then gives strong support for some new exotic mechanism like,
most plausible, the temporary formation of a deconfined and strangeness
saturated new state of matter.

Although intriguing, after all this may not be the correct interpretation
of the observed antihyperon yields:
In the following we
elaborate on our recent idea of antihyperon
production by multi-mesonic reactions like
$n_1\pi + n_2 K \rightarrow \bar{Y}+p $
corresponding to the inverse of the strong binary
baryon-antibaryon annihilation process \cite{GL00}.
The latter process, on the other hand, dictates the timescale
of how fast the antihyperon densities do approach local
chemical equilibrium with the pions, nucleons and kaons.
This timescale is thus to a good approximation proportional to the
inverse of the baryon density.
Adopting an initial baryon density of approximately 1--2 times
normal nuclear matter density $\rho_0 $ for the initial and thermalized
hadronic fireball, the antihyperons will equilibrate
on a timescale of 1--3 fm/c! This timescale competes
with the expansion timescale of the late hadronic fireball, which
is in the same range or larger. In any case it becomes plausibel
that these multimesonic, hadronic reactions, contrary to the binary
reactions (\ref{sprodb}), can explain  most conveniently a sufficiently
fast equilibration {\it before} the (so called) chemical freeze-out occurs at
the parameters given by the thermal model analyses (, and where
the baryon density has dropped to around 0.5 - 0.75 $\rho_0 $).
One can always argue that before this point in time a new state of matter
might have occured as the energy density becomes close to or above 1 GeV/fm$^3$
\cite{Heinz}. (This is the value what lattice QCD at present
estimates for the critical energy density, where at equilibrium
and zero net baryon density
the transition to a deconfined state should occur.)
Our interpretation does rest on the (conservative) view that before the
chemical freeze-out already a hadronic system has established.
Whether even before this stage a deconfined state or a
non-equilibrium stage of hadronic string-like
excitations had existed is then still a matter of debate, but
it is not the present issue for explaining the chemical saturation
of the antihyperons at chemical freeze-out.

Before we detail this mechanism of antihyperon production and also
comment on a few of the necessary assumptions
(and potential reservations \cite{Heinz}), we first want to briefly sketch
in the next section a couple of interesting conclusions
on overall strangeness production (i.e. the most dominant kaons and
$\Lambda $s)
and equilibration obtained within a microscopic hadronic transport model
\cite{Ge98,BCGEMS00}.

\section{Strangeness production and equilibration}

\begin{figure}[htb]
\vspace*{1cm}
\begin{center}
\begin{minipage}{10cm}
\epsfxsize=8cm
\epsfbox{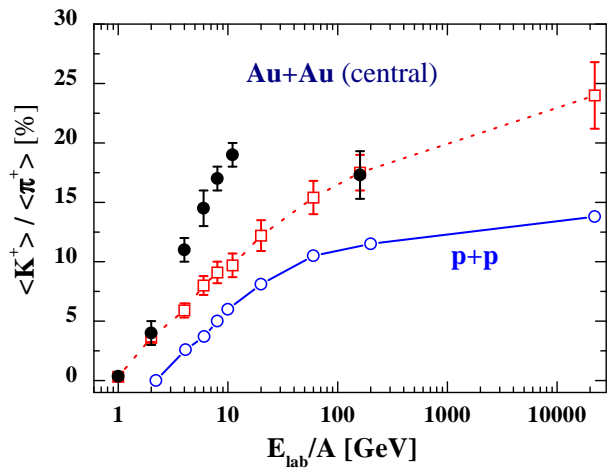}
\end{minipage}
\end{center}
\caption[]{Calculated $K^+/\pi^+$-ratio around midrapidity
for central Au+Au reactions (open squares) from SIS to RHIC energies
in comparison to experimental data. For visualization of the
the collective strangeness enhancement the corresponding ratio for
elementary p+p collisions (open circcles) is also depicted.
This plot has been taken from \cite{Ge98a,Ca99}.}
\label{fig:Jochen}
\end{figure}

\begin{figure}[htb]
\begin{center}
\begin{minipage}{125mm}
\epsfysize=6cm
\epsfbox{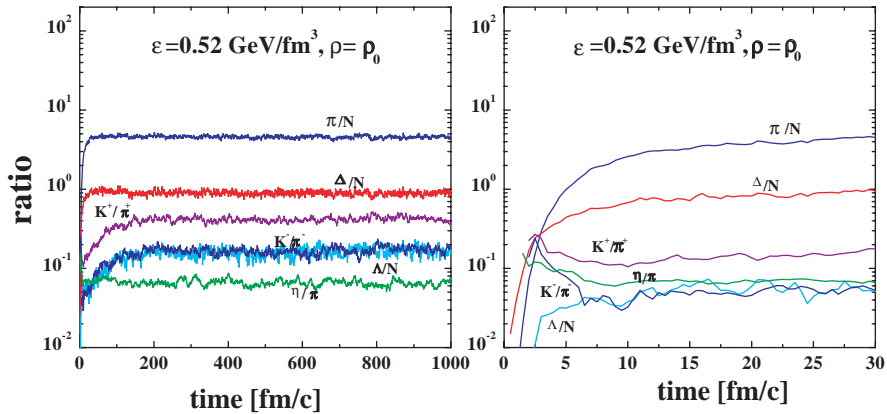}
\end{minipage}
\end{center}
\caption[]{Time evolution
towards stationary equilibrium values for
various particle ratios in an `infinite matter' calculation \cite{BCGEMS00}.
The box is prepared with a baryon density $\rho_B=\rho_0$ and an
energy density of $\epsilon = 0.52 \,$ GeV/fm$^3$.
The $K^+/\pi^+$ ratio needs about 50-100 fm/c to approach to its
stationary equilibrium
value.
The left panel shows the time scale up to
1000 fm/c, whereas the right panel demonstrates the initial stage.
For the initialized non-equilibrium situation the $K^+/\pi^+$-ratio
starts via the decay of the early (primary, secondary and ternary)
string excitations already at a large value moderately close to its later
equilibrium value.}
\label{fig:Elena2}
\end{figure}

As outlined in the introduction the possible
strangeness enhancement in heavy ion collisions
has been one of the driving motivation for the experimental study
of strange particle production.
Since a relative enhancement is observed already in hadron-hadron
collisions  for increasing energy (see Fig.~\ref{fig:Jochen}), which is
certainly not due to any macroscopic or bulk effects, the
to be measured strangeness should be compared
relative to p+p collisions at the same energy.
The arguments for enhanced strangeness
production via the QGP should generally apply already for the most dominant
strange particles, the kaons, as their chemical equilibration time
in a hadronic fireball has been estimated to be
$\approx 100$ fm/c \cite{KMR86} (see also fig. \ref{fig:KMR} and
fig. \ref{fig:Elena2}). On the other hand, it was also argued
that a factor of 2-3 enhancement in the $K/\pi $-ratio relative to the one
obtained in p+p collisons
can only be seen as an indirect signal for QGP creation \cite{KM86}.
Moreover, as we will now summarize, nonequilibrium inelastic hadronic reactions
can explain to a very good extent the overall strangeness production
seen experimentally \cite{Ge98,Ma89}.

In a recent systematic study we had investigated the properties of $K^+ $, $K^-$ and
$\Lambda $ particles in nuclear reactions from SIS to CERN-SPS energies \cite{Ge98}
within the microscopic hadron-string transport approach HSD (for details describing
the transport algorithm see \cite{CB99}). An important ingredient has been
the implementation of
the elementary cross sections for strangeness production in baryon-baryon,
baryon-meson and meson-meson channels. An enhancement of the scaled
kaon yield due to hadronic rescattering both with increasing system size and energy
was found. This is expected within any hadronic model if the kaons
and other particles do not feel any attractive potentials.
After the {\em primary} string fragmentation
of intrinsic p-p--collisions the hadronic fireball starts with a
$K^+/\pi^+$ ratio still far below chemical equilibrium with $\approx 6 - 8 \% $
at AGS to SPS energies before the hadronic rescattering starts.
As the average kinetic energy and the particle density increases monotonically
with incoming kinetic energy of the projectile while the lifetime of the
fireball increases with the system size, a smooth and continous
enhancement is expected in a hadronic description by these effects.
The outcome for the most dominant strange particles, the
$K^+$-mesons, is summarized in	fig. \ref{fig:Jochen}.


%

We want to emphasize that the secondary
(meson-baryon) and ternary (meson-meson)
induced string-like interactions do contribute significantly
to the total strange particle production, particular for reactions
at SPS energies.
Via these channels about the same number of strange and anti-strange quarks
is produced as in the primary p+p collisions. This then explains the
factor 1.75 as the relative enhancement compared to p+p
(compare fig. \ref{fig:Jochen}).
Hence, the major amount
of produced strange particles (kaons, antikaons and $\Lambda $s) at SPS-energies
can be understood in terms of early and still energetic,
non-equilibrium interactions.
On the other hand, at the lower AGS energies, the relative enhancement
factor of $\approx 3$ can not be fully explained within the cascade
type calculations \cite{Ge98}. This might indicate some new physics involved
for the primary $s\bar{s}$ production mechanism:
Including meson potentials can help to reasonably
understand the production of $K^+$ and especially $K^-$ mesons
at lower SIS energies,
yet some smaller, but still significant underestimation at AGS energies
does persist \cite{CB99,Ca99}.

Only for a system close to thermal equilibrium, as was assumed
in the early calculations \cite{KMR86}, the
overall strangeness production rates (for kaons and $\Lambda $s)
are substantially suppressed due to the high thresholds.
As pointed out above and also in a very recent study \cite{BCGEMS00}
this is due to the oversimplified initial conditions.
In \cite{BCGEMS00} `infinite' hadronic matter is simulated
within a cubic box, starting
with a nonequilbrium initial configuration in momentum space
which does somehow resemble the initial or early intermediate situation
in a true heavy ion collision.
One particular microscopic simulation
towards equilibrium
is depicted in fig. \ref{fig:Elena2}.
As one can see really chemical equilibrium for the kaons and antikaons
is approached only at $\approx 50$ fm/c at
the given energy and baryon density (which both are higher than the ones
calculated from the chemical freeze-out point \cite{BMS96}).
This is in accordance
with the early calculations \cite{KMR86}.
On the other hand one also sees that eg
the $K^+/\pi^+$-ratio $\approx 0.15$
starts via the decay of the early
string excitations already at a quite large value and then stays
rather constant in time.
As elaborated above, in a simulation of a true heavy ion collision
strangeness is produced in the very early stage and
these early produced strange/antistrange
quarks then suffice to
explain the majority of strange particles (kaons, antikaons and $\Lambda $s)
at SPS energies.

In addition, it was also shown in \cite{BCGEMS00} that local kinetic equilibrium
is reached on a sufficient fast timescale by the multiple processes
of subsequent string fragmentation. The string excitations do provide a
very efficient mechanism to produce transversal energy. In summary,
the microscopic transport
calculations do support qualitatively the idea that there exists
a regime in time during the heavy ion collision where thermal and chemical
equilibrium among the various hadronic particles should be (locally) realised.

\section{Antihyperon production by kaons and pions}

We now repeat and detail on our previous idea on antihyperon
production \cite{GL00}: Not subsequent binary hadronic reactions
of type (\ref{sprodb})
but in fact multi-pionic and kaonic interactions
in a thermalized hadronic gas
lead to a very fast chemical equilibration of the antihyperon
degrees of freedom.
For this one has to look first on the following annihilation reactions
similar to the standard baryon annihilation $\bar{p} + p \rightarrow
n \, \pi $, but now
involving one antihyperon and then do apply rigorously the concept
of detailed balance:
\begin{eqnarray}
\bar{\Lambda } + N  & \leftrightarrow & {n}_{\bar{ \Lambda }}\,  \pi  + K  \nonumber
\\[2mm]
\bar{\Xi } + N  & \leftrightarrow & {n}_{\bar{ \Xi }}\,   \pi  + 2  K
\nonumber
\\[2mm]
\bar{\Omega } + N  & \leftrightarrow & {n}_{\bar{ \Omega }}\,   \pi  + 3  K \nonumber
\end{eqnarray}
or, in shorthand notation,
\begin{equation}
\label{antihyp1}
\bar{Y} + N \, \leftrightarrow \, {n}  \pi + n_Y  K
\, \, \, .
\end{equation}
$n_Y$ counts
the number of anti-strange quarks within the antihyperon $\bar{Y}$.
${n}+ n_Y$ is expected to be around $\approx 5-7 $.
The reactions (\ref{antihyp1}) are all exothermic.
It is also plausible to assume that the
annihilation cross sections are approximately the same like for $N\bar{p}$
at the same relative momenta. Hence, in the relevant regime
of a thermal hadronic gas with temperatures of $T\approx 120 - 200 $ MeV
one has $\sigma _{p \bar{Y}\rightarrow n  \pi + n_Y K}
\approx \sigma _{p\bar{p} \rightarrow n \pi}  \approx 50 $ mb, which
is indeed a large cross section.

\begin{figure}[htb]
\begin{center}
\begin{minipage}{9cm}
\epsfysize=6cm
\epsfbox{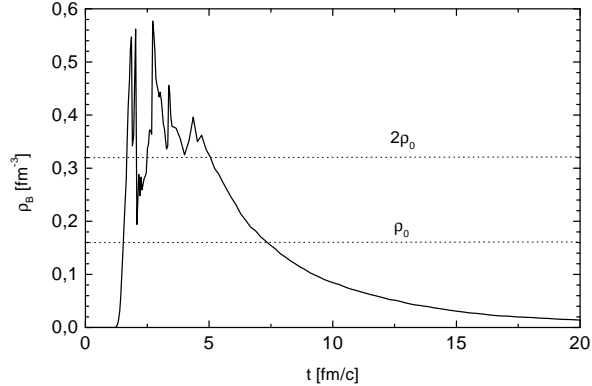}
\end{minipage}
\end{center}
\caption[]{Time evolution of the (average) net baryon density
for midrapidity $|\Delta Y| \leq 1$ and central Pb+Pb-collision.
Here the amount of baryon number residing still in string-like
excitations is explicitely discarded.
String-like excitations have disappeared after 4.7 fm/c, so that
from this time on a pure hadronic fireball develops and expands.
Its initial net baryon-density starts slightly above $\rho_B=2\rho_0$.
}
\label{fig:WC}
\end{figure}

The above reactions (\ref{antihyp1}) do effectively lead
to the following master equation for the respectively considered
antihyperon density
within a hadronic gas:
\begin{equation}
\label{mastera}
\frac{d}{dt} \rho _{\bar{Y}} \, =\,   -
\expl \sigma _{\bar{Y}N} v _{\bar{Y}N} \expr
\left\{
\rho _{\bar{Y}} \rho_N \,  \vphantom{\sum_{n}}
-  \, \sum_{{n}}
{\cal R}_{(n,n_Y)}(T,\mu_B,\mu_s) (\rho _\pi)^{{n}} (\rho _K )^{n_Y}
\right\} \, \, \, .
\end{equation}
The `back-reactions' of several effectively coalescing
pions and kaons are incorporated by the `mass-law' factor
\begin{equation}
{\cal R}_{(n,n_Y)}(T,\mu_B,\mu_s) \, = \,
\frac{ \rho _{\bar{Y}}^{eq.} \rho ^{eq.}_N }
{(\rho ^{eq.}_\pi)^{{n}} (\rho ^{eq.}_K )^{n_Y}} \, p_n \, \, \, .
\end{equation}
Here $p_n$ states the (unknown) relative probability
of the reaction (\ref{antihyp1}) to decay into a specific number $n$ of pions
with $\sum_n p_n =1$. ${\cal R }$ has a clear physical origin as
it is responsible to {\em assure detailed balance} in the competition
between
the annihilation process and the various contributing
multi-mesonic `back reactions'.
${\cal R}$ then depends only on the temperature and the baryon and strange quark chemical
potentials.
$\expl \sigma _{\bar{Y}N} v _{\bar{Y}N} \expr$ denotes the thermally averaged
cross section. We take $N$ as synonym for any baryonic particle, most
dominantly the nucleons and $\Delta $-excitations.
Furthermore,
$\Gamma  _{\bar{Y}} \equiv
\expl \sigma _{\bar{Y}N} v _{\bar{Y}N} \expr \rho_B $
gives the effective
annihilation rate of the respective antihyperon specie on a baryon.
Assuming that
the pions, baryons and kaons stay in thermal and chemical
equilibrium, the master equation then becomes simply
\begin{equation}
\label{masterd}
\frac{d}{dt} \rho _{\bar{Y}} \,  = \,    - \,
\Gamma  _{\bar{Y}}
\left\{
\rho _{\bar{Y}} \, -  \,
\rho ^{eq }_{\bar{Y}}
\right\} \, \, \, .
\end{equation}
It should become clear by now that indeed the mean annihilation rate
yields the characteristic inverse time to drive the antihyperon densities
to their chemical equilibrium values, i.e. it corresponds to the inverse of the
characteristic chemical equilibration time. So how large is it?

At the onset of thermalization
and chemical equilibration for all other degrees of freedom
in the hadronic fireball
the baryon density might still be rather large and could exceed two
times normal nuclear matter density \cite{So96,Cass}. In fig. \ref{fig:WC}
we have depicted the net baryon density as a function of time
at a space region for particles at midrapidity obtained within a
microscopic transport model \cite{Cass}. The figure illustrates
that a pure hadronic fireball (without any
string-like excitations) at two times baryon density
has established during the ongoing (longitudinal) expansion.
It is interesting to note that the chemical freeze-out `point'
with the parameters calculated in \cite{BMS96,CR00,KM00} takes place
at a value of $0.5-0.75 \, \rho_0$.
This would correspond to a time of 8-10 fm/c
in fig. \ref{fig:WC}. Taking now for the average baryon density
evolving shortly before the chemical freeze-out point
$ <\rho_B > \approx 1-2 \rho_0 $ and employing the above estimate
for the antihyperon annihilation cross scetion, i.e.
$\sigma _{p \bar{Y}\rightarrow n  \pi + n_Y K} \approx 50 $ mb, one
has for the chemical equilibration time
of antihyperonic particles the striking number
\begin{equation}
(\Gamma _{\bar{Y}})^{(-1)}  \, = \,
\frac{1}{\expl \sigma _{N \bar{Y}\rightarrow n  \pi + n_Y K}
v _{\bar{Y}N} \expr <\rho_B > }
\, \approx \,  1 - 3 \, \mbox{fm/c}
\, \, \, .
\label{taueq}
\end{equation}
This is a very fast process (!) and lies
below the typical fireball lifetime of $10$ fm/c.
(Indeed, microscopic calculations within (U)RQMD have shown
that antibaryon annihilation takes place with considerable rate
and that the overall anti-baryon yield can be hardly
described within the standard transport approaches
exactly because of the large
annihilation cross section \cite{S95}.)
Antihyperons are forced rather immediately to local
chemical equilibrium together with the pions, kaons and nucleons
by the `back reactions'!
One has to be a little bit more precise \cite{Heinz}:
What actually has to be compared is the timescale of
how fast the fireball does expand or, refering
to the rate (\ref{taueq}), of how fast the baryon density does
drop. From fig. \ref{fig:WC} one finds that this timescale
is 3-4 fm/c. This is a reasonable expectation.
Hence, there is no need for any `exotic' explanation (like
eg the temporal existence of a potential QGP
saturated in strangeness) to account for the thermally
and chemically equilibrated particle number of antihyperons
observed at the chemical freeze-out point.
In fact, beyond that `point' (which, of course, is actually
some continous regime where inelastic decoupling occurs) with already
a moderately low baryon density (and correspondingly
low pion and kaon densities) it will be that the multi-mesonic
creation process
becomes more and more ineffective. This would then also
explain the claer `position' of the chemical freeze-out point
for the antihyperons.

To be more quantitative some
explicite coupled master eqautions for an expanding system have to be
considered. Such work is in progress \cite{GLnew}.
In addition one can also study at which point on average
the antihyperon degrees of freedom kinetically do decouple
(thermal freeze-out). The decoupling does depend
probably and most simply on the explicit (and unknown)
parametrisation of the elastic cross section of the antihyperons with the
pions. This has been pointed out already by Hecke et al \cite{So98} when
adressing the fact that the experimentally deduced effective inverse
slopes $`T_{Y}'$ of the (anti-)hyperon spectra and especially of the
multi-strange $\Omega $-spectrum
do not follow the linear increasing trend in mass.

\section{Summary, conclusions and outlook}

To summarize,
the multi-mesonic source of production of antihyperons is a consequence
of detailed balance and, as the rate $\Gamma_{\bar{Y}}$ is
indeed very large, this is the by far most dominant source
compared to any binary production channel (\ref{sprodb}).
This, as we believe, is a remarkable observation
as it clearly demonstrates
the importance of hadronic multi-particle channels.
At the moment such `back-reactions' cannot be handled
within the present transport codes and some clever strategy
has to be invented. This could be a nice exercise for the future.
Nonetheless, as we have shown, there exists a simple non-exotic
mechanism $^{c,d}$ for explaining the $\bar Y$ abundancies
in a purely hadronic scenario.

One might be tempted to ask whether a similar reasoning also applies
for the multi-strange hyperons (the $\Xi $ and the $\Omega $) for
which also some significant enhancement has been reported.
The answer is `no'. The equilibration rate here would be governed by the
density of antibaryons and is thus too low, or putting it differently,
the equilibrium density of multi-strange hyperons is much higher
than the one of antihyperons. It might be that only more exotic
microscopic processes (or potentially only the celebrated deconfined
state of matter) can explain the enhancement \cite{SBBBZSG99}.

The mechanism to work out
for the antihyperonic degrees of freedom is based
on two rather moderate assumptions:
(I) The thermally averaged annihilation cross section
for antihyperons colliding with a nucleon, i.e. $\bar{Y} + N $,
is roughly as large as the measured one for $\bar{p} + p$
or $\bar{p} + n$.
(II) At the onset for the equilibration of the antihyperons
one has to assume a hadronic fireball with still a moderate
baryonic density and where the pions together with the
nucleons {\em and} the kaons are assumed to be
nearly in chemical equilibrium.
As discussed in the second section, the
abundant and early production of
kaons and antikaons can reasonably be accounted for by
hadronic transport models.
If, as presented in some of the thermal models,
a strangeness suppression factor $\gamma_s $ for each unit of strangeness
is introduced \cite{CR00}, one then finds for the
stationary point of the master equation \cite{GL00}
\begin{equation}
\label{semieq}
\rho _{\bar{Y}}^{eq} \rightarrow
\rho _{\bar{Y}} =
(\gamma_s )^{n_Y} \rho _{\bar{Y}}^{eq.}
\, \, \, ,
\end{equation}
which is consistent
with the employed phenomenological prescription \cite{CR00}.

There is also a clear hint at AGS energies
of enhanced anti-$\Lambda $ production:
On the one hand the E859 Collaboration has measured the $\bar{\Lambda }/\bar{p} $
ratio in Si+Au at 14.6 AGeV and had reported a large value
$\bar{\Lambda }/\bar{p} = 2.9 \pm 0.9 \pm 0.5 $ for some central
rapidity window. On the other hand
such a value has also been discovered, albeit for low transverse momentum,
by the E864 Collaboration for the most central collisions
\cite{E859}.
According to the thermal models the deduced temperatures
at the AGS-energies are lower
and the obtained baryon densities are even higher.
Our argument should thus perfectly apply.
Measurements of antihyperon production
could also be done at possible future heavy ion facilities
at GSI working then at much higher bombarding energies
comparable or exceeding AGS energies. This would
be a very interesting opportunity to unreveal the here proposed mechanism.

\section*{Acknowledgements}
The work presented had been done in various collaborations
with E.~Bratkovskaya, W.~Cassing, J.~Geiss, S.~Leupold and U.~Mosel.
I am in particular indebted to W.~Cassing for providing fig.
\ref{fig:WC} and to S.~Leupold, with whome the idea
of multimesonic antihyperon production has been developed and further
pursued.
I also want to thank U. Heinz for the many interesting,
albeit controversial discussions and J. Nagle for
discussing the status of the $\bar{Y}$-production at AGS.
This work has been supported by BMBF and GSI Darmstadt.

\section*{Notes}
\begin{notes}
\item[a]
Invited talk at the Symposium on `Fundamental Issues in Elementary Matter',
241. WE-Heraeus-Seminar, 25-29 September, Bad Honnef, Germany.
\item[b]
E-mail: Carsten.Greiner@theo.physik.uni-giessen.de
\item[c]
Our idea has been
triggered by a recent work \cite{RS00} (but see also next footnote $^d$)
which dealt with the question of how antiprotons might maintain
nearly perfect chemical equilibrium until so called thermal freezeout
at temperatures $T\approx 110 $ MeV much lower than at chemical freeze-out.
However, the baryon densities are typically much lower between these
two stages of chemical freeze-out to thermal freeze-out
($\rho_B \sim 0.5-0.05 \rho_0$, compare also fig.~\ref{fig:WC}),
so that the multipionic reactions
$ n \pi \leftrightarrow  N + \bar{p} $ becomes less effective while
competing against the rather rapid expansion and dilution.
Our intention, on the other hand, has been to explain
qualitatively via the adressed multihadronic channels the production
of antihyperons before and at the so called point of chemical freeze-out.
\item[d]
The here discussed
multi-mesonic channels for producing antihyperons
are not considered for the first time:
In fact they had been taken into account
in the set of master equations for the
strange hadronic particle densities by Koch et al \cite{KMR86}.
The now mysterious question is then why the authors had not
come at that time to our present conclusion?
Much to the contrary they put forward the
agenda for the antihyperons as a clear
signature of a QGP. Looking at Fig.~B3 in \cite{KMR86} they
have only considered the annihilation cross section $
\sigma_{p\bar{p}\rightarrow 5 \pi} \approx 10$ mb,
which is a factor of 5 or so smaller than the total
annihilation cross section.
Still, inspecting fig. \ref{fig:KMR}, their equilibration
rate of the antihyperons is even then still two to three orders of
magnitude too small!
\end{notes}

\vfill\eject
\end{document}